# Advanced $Cd_{1-x}Mn_xTe:Fe^{2+}$ semiconductor crystals with linear *sin-band* redshift of absorption and emission spectra

Sergei V. Naydenov[*], Oleksii K. Kapustnyk, Igor M. Pritula[**], Nazar O. Kovalenko, Igor S. Terzin, Dmitro S. Sofronov, Pavel V. Mateichenko

Institute for Single Crystals of the National Academy of Sciences of Ukraine,
60 Nauky Avenue, Kharkiv 61072, Ukraine

**Abstract**

Doped semiconductor crystals of solid solution $Cd_{1-x}Mn_xTe:Fe^{2+}$ were grown by the high-pressure Bridgman method covering the range of its existence as a zinc blende crystal structure ($0 < x < 0.77$). The concentration of $Fe^{2+}$ impurities was approximately $10^{-3}$ wt.% in all studied samples. The structural and optical properties of the crystals were investigated, including the case of high concentrations of $x > 0.45$. The correlations between the composition of solid solution crystals of $Cd_{1-x}Mn_xTe:Fe^{2+}$, band gap, lattice period and the maxima positions of the $Fe^{2+}$ active ion absorption and emission spectra were found. A new theoretical model based on the principle of additivity for solid solution semiconductor materials has been used for explaining the long-wavelength "redshift" of absorption and luminescence bands in the spectra of $Cd_{1-x}Mn_xTe:Fe^{2+}$ crystals with increasing solid solution concentration. The obtained results can be used to predict the lasing range for $Cd_{1-x}Mn_xTe:Fe^{2+}$ crystal media (in all possible Mn concentrations), which is potentially the longest wavelength active material for mid-IR lasing.

## 1. Introduction

Crystalline materials based on semiconductor compounds of Groups II and VI doped with transition metal ions ($A^{II}B^{VI}:TM^{2+}$) are promising laser media for generation in the near- and mid-IR range [1, 2]. Undoped semiconductor crystals such as CdTe and its solid solutions CdZnTe, CdMnTe and others are also famous detector materials for the registration of different types of ionizing radiation.

Various applications require developing IR laser media with broad tunable generation bands shifted toward long wavelengths (by up to 5-6 µm). New possibilities in comparison with binary compounds are presented by solid solutions of complex chalcogenides in which the main cation in the crystal lattice is substituted with a lighter atom. Obtaining solid solutions is an effective way to modify the optical and luminescent properties of chalcogenide-based active materials. Analysis of the experimental data shows that one of the most promising materials for shifting the lasing band toward longer wavelengths is $Cd_{1-x}Mn_xTe:Fe^{2+}$ [3-8].

Crystals of solid solution $Cd_{1-x}Mn_xTe$ are prospective materials for tunable laser sources emitting in the middle infrared range as well as for room-temperature radiation detectors. At Mn concentrations of $x \sim 0.05$, these crystals may be used as radiation detectors [9-13]. At higher concentrations of Mn, from $x = 0.1$ up to $x = 0.76$, the crystals doped with iron (active impurity) are of interest as active elements of tuneable mid-IR lasers in the 3-5 µm region.

---

CdMnTe solid solutions have several advantages compared with other $A^{II}B^{VI}$ compounds, including (i) better lattice strengthening and mechanical stability; (ii) wide bandgap tuning in the range of 1.7-2.2 eV due to the strong compositional influence of Mn; (iii) high resistivity and good electron transport properties; and (iv) near-unity segregation coefficient of Mn (this property has a positive effect on the homogeneity of electrical and optical properties of grown crystals). The host crystal field forms such an energy spectrum of iron dopant ions that the resulting mid-IR band of luminescence and the optical gain are exhibited at the longest wavelengths among all known host $A^{II}B^{VI}$ binary compounds and their solid solutions.

Thus, there is great interest in obtaining a series of solid solutions of $Cd_{1-x}Mn_xTe:Fe^{2+}$ and studying their optical properties to obtain an understanding of the correlation between the structural parameters of the crystalline host and the optical properties of the active ion.

## 2. Experimental

Crystals of $Cd_{1-x}Mn_xTe:Fe^{2+}$ were grown by the high-pressure Bridgman method in a glassy carbon crucible under argon pressure with different mole fractions of Mn. We grew crystals throughout the whole range of concentrations between x = 0.09 and x = 0.77, in which this compound can exist in the zinc blende structure. The upper concentration is limited by the existence of a solid solution [12]. The charges were prepared from high purity metallic Cd, Mn, and Te and were stoichiometrically weighed for each composition. The iron impurity concentration was on the order of $10^{18}$ cm$^{-3}$. It was the same at ~$10^{-3}$ wt.% in all studied samples.

The growth setup was designed at our department for AIIBVI crystal growth, and it has a dedicated automated growth control system. This ensures a reduction in concentration fluctuations throughout the volume of a crystal ingot. The growth mode was adjusted by a temperature gradient of approximately 30°C/cm and a crucible travel speed of 0.5-0.6 mm/h. It was optimized experimentally by the surface quality of the cleavage plane of the grown crystals. It has been established that at higher growth velocities, the concentration of twins is much higher. The Ar pressure in the growth chamber was approximately 20 bar. We obtained nearly single crystal ingots approximately 30 mm in diameter and 50-70 mm in length comprising crystallites several cubic cm in volume (see Fig. 1).

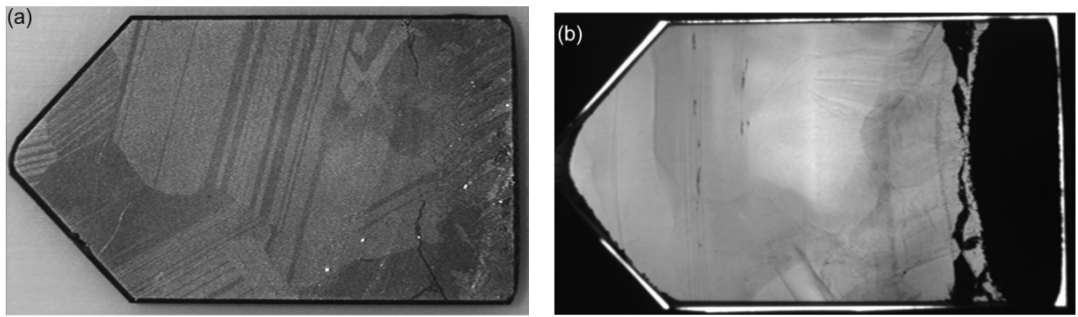

**Fig. 1.** Images of typical crystal $Cd_{1-x}Mn_xTe:Fe^{2+}$ longitudinal slices made in (a) reflected visible IR and (b) transmitted IR light.

The elemental composition and SEM images were defined by using a JSM-6390LV scanning electron microscope (JEOL Ltd., Japan) with an X-ray energy-dispersive analysis system X-Max 50 (OXFORD Instruments Analytical, Great Britain). Optical spectrophotometry for visible, near and mid-infrared regions was performed by a PerkinElmer Lambda 35 spectrophotometer (USA) in the visible range and a PerkinElmer Spectrum One spectrophotometer (USA) in the range of 1.28-25 μm. Absorption and emission spectra in the mid-IR and visible ranges of the optical spectrum were studied.

## 3. Theory

Transition metal-doped crystals of $A^{II}B^{VI}$ compounds are effective active materials for tunable mid-infrared lasers. Lasing of $TM^{2+}$ ions was carried out by a quasi-four-level scheme [1].

The differences between the ion radii and electronegativities of $TM^{2+}$ and crystal-lattice cations (Zn, Cd, Hg, etc.) usually do not exceed 10-15% and 0.4-0.6 Poling units, respectively. Therefore, the $TM^{2+}$ ion can, according to the Goldschmidt rule, isomorphously substitute for a crystal-lattice cation. As a result, the "central" ion occurs in the octahedral or tetrahedral environment of neighboring chalcogen anions (S, Se, Te). The energy levels of this impurity ion fall deep into the semiconductor bandgap [14, 15]. The luminescence of $TM^{2+}$ ions in the IR range has an "intra-center" character. The processes of charge exchange between $TM^{2+}$ ions under irradiation with visible and UV light or due to impact ionization by free electrons in a strong (several kilovolts) electric field and related recombination (short-wavelength) luminescence will not be considered below.

Optical absorption and tuning band of $Fe^{2+}$ ion in $Cd_{1-x}Mn_xTe:Fe^{2+}$ solid solutions and its spectral location is defined by internal energy gap between ground state $^5E$ and excited state $^5T_2$. We name this splitting energy $\Delta E^*$. It is one of the main parameters of the theory. In substitutional solid solutions, the impurity ion can replace any of the two kinds of cations (rather than a single one). Since the splitting energy $\Delta E^*$ of the $TM^{2+}$ ion depends on the local environment, it cannot be expressed by the famous Bethe's formula (see, e.g., [16]). To relate this energy value to the parameters of the semiconductor material, let us naturally assume that, on the passage from simple compounds to their solid solution, a physical quantity should vary (to the first approximation) in proportion to the concentrations of components in the mixed system. In a particular case, this principle as applied to lattice parameter $d(x)$ corresponds to the well-known Vegard law $d(x) = d_1 + (d_2 - d_1)x$, which is quite valid for the materials under consideration. In application to the energy gap $\Delta E^*$ between states of the $TM^{2+}$ ion, this principle of *additivity* implies that [17]

$$\Delta E^*_{sol}(x) = (1 - x)\, \Delta E^*_1 + x\, \Delta E^*_2, \qquad (1)$$

where $\Delta E^*_{sol}$, $\Delta E^*_1$ and $\Delta E^*_2$ are splitting energies for $Fe^{2+}$ ions placed in the matrix of solid solution $Cd_{1-x}Mn_xTe:Fe^{2+}$ or its "pure" or "undiluted" binary compounds $CdTe:Fe^{2+}$ and $MnTe:Fe^{2+}$, respectively. Note that, in the framework of this approach, it is also possible to take into account the effect of temperature and Jan-Teller's effect by adding to Eq. (1) the corresponding terms not considered here. A theoretical analysis of energy levels and energy transitions shows that from Eq. (1), it follows

$$\varepsilon(x) \approx \varepsilon_1 - \delta_{12}\, x = \varepsilon_2 + \delta_{12}\, x, \qquad (2)$$

where $\varepsilon(x)$ is the energy of the photon absorbed or emitted during the intra-center transition in the $Fe^{2+}$ ion in the matrix of solid solution with concentration $x$. Energies $\varepsilon_1$ or $\varepsilon_2$ correspond to the case $x=0$ or $x=1$, i.e., to the energy of the photon absorbed or emitted in binary crystals $CdTe:Fe^{2+}$ or $MnTe:Fe^{2+}$, respectively. The energy difference $\delta_{12} = \varepsilon_1 - \varepsilon_2$ does not depend on the concentration of the solid solution.

Let us consider the change in the photon energy $\varepsilon(x)$ due to the change in solid solution composition $\Delta x$. The wavelength shift $\Delta\lambda$ is related to the photon energy change $\Delta\varepsilon$ as

$$\Delta\lambda/\lambda = -\Delta\varepsilon/\varepsilon. \qquad (3)$$

In the case of small increments, $\Delta\lambda \ll \lambda$ or $\Delta\varepsilon \ll \varepsilon$, it follows from Eqs. (1)-(2) that

$$\Delta\lambda_{ab} \approx K_{ab}\, \Delta x; \qquad (4)$$

$$\Delta\lambda_{em} \approx K_{em}\, \Delta x; \qquad (5)$$

$$K_{ab} \sim \lambda_{ab}\, K;\ K_{em} \sim \lambda_{em}\, K; \qquad (6)$$

$$K = \delta_{12}/\varepsilon, \qquad (7)$$

where wavelength $\lambda_{ab}$ and $\lambda_{em}$ are linked to the absorption and emission (luminescence) spectra. Here, coefficient $K$ is the (dimensionless) microscopic coefficient of wavelength shift, and we suggest that it is small ordinarily, i.e., $K \ll 1$. Eqs. (4)-(5) describe the law of linear shift of the maximum of absorption or emission band depending on change $\Delta x$ in the concentration of solid solution. The $\lambda$ values correspond to maxima in the spectra of absorption $\lambda_{ab}$ or emission $\lambda_{em}$ of the "undiluted" binary compound.

Eqs. (4)-(7) used for the $Cd_{1-x}Mn_xTe:Fe^{2+}$ system can be presented in the form

$$\frac{\Delta\lambda}{\lambda} \sim K\Delta x, \qquad (8)$$

$$K = 1 - \frac{\Delta E^*_{MnTe}(Fe^{2+})}{\Delta E^*_{CdTe}(Fe^{2+})}, \qquad (9)$$

where $\Delta E^*$ is the Stark splitting energy of the central $Fe^{2+}$ ion in the crystal field of the corresponding binary components. Theoretical estimates of parameters $\delta_{12}$ and $K$ were calculated for $Cd_{1-x}Mn_xTe:Fe^{2+}$ ternary crystals. Table 1 presents the corresponding microscopic parameters of the "undiluted" binary components that were used for estimation of $\Delta E^*$. We obtained values of $\delta_{12} \sim 1.2 \times 10^{-2} > 0$ and $K \sim 6.3 \times 10^{-2}$. The sign of $K > 0$ is positive, so that the bands shift toward longer wavelengths. The typical orders of magnitudes for band-shift parameters are $\varepsilon \propto 10^{-1}$ eV, $\delta \propto 10^{-2}$ eV, and $K \propto 10^{-1}$.

**Table 1.** Parameters of the undiluted binary compounds $CdTe:Fe^{2+}$ and $MnTe:Fe^{2+}$ of solid solution $Cd_{1-x}Mn_xTe:Fe^{2+}$. The radius of an external valence $d$-electron in the $Fe^{2+}$-ion was roughly approximated as the impurity atom radius.

| Parameter | CdTe | MnTe |
|---|---|---|
| $e$, dielectric permittivity, rel. units | 12.82 | 19.30 |
| $r$, ion radius of cation, Å | 0.92 | 0.80 |
| $d$, lattice parameter, Å | 6.48 | 6.33 |
| $r_1$, radius of d-electron of $Fe^{2+}$, Å | 1.26 | 1.26 |

The character of the spectral shift is determined by the sign of $\delta_{12}$, whereby the long wavelength and short wavelength correspond to $\delta_{12} > 0$ and $\delta_{12} < 0$, respectively. In the general case, the value of the shift coefficient $K_\alpha$ depends on the structure of the energy band spectrum of a particular $TM^{2+}$ ion in the matrix of the solid solution. For each separate $\alpha$-th band in a complex spectrum of absorption or emission, this coefficient can take different (albeit usually close) values. The main contribution to the value of $K_\alpha \sim K$ is related to Stark splitting, i.e., to the $\Delta E^*(x)$ value. From Eqs. (4)-(5), it follows that the clear relation

$$\Delta\lambda_{ab}/\lambda_{ab} \sim \Delta\lambda_{em}/\lambda_{em}, \quad (10)$$

that may be tested experimentally. Eq. (10) means that under the changing of solid solution composition $\Delta x$, the shift of absorption and emission spectra occurs by the *sin-band* way. Both maxima of absorption and emission spectra should have the same direction of the shift to longer (red-shift) or shorter (violet-shift) wavelengths.

## 4. Results and discussion

We studied the composition and structural properties of the grown crystals. XRD analysis of the composition of all grown crystals suggests substitution of the solid solution with a mixed cation sublattice Cd (Mn). The typical chemical element distribution along the ingot of CdMnTe:$Fe^{2+}$ is shown in Fig. 2. All samples have satisfactory axial and radial homogeneity along and across the ingots. Crystal growth by the high-pressure Bridgman method allows large crystals with homogeneous compositions to be obtained. The ingots contain large single crystalline blocks.

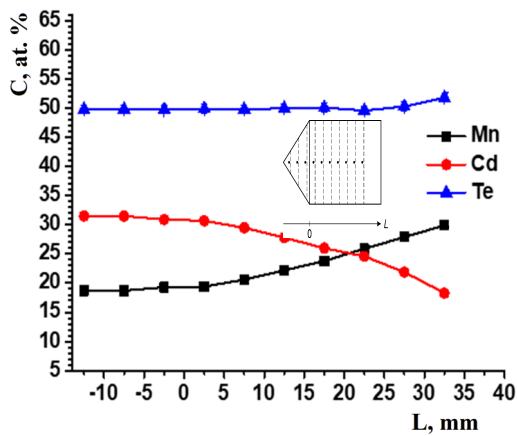

**Fig. 2.** Chemical element distribution of the solid solution of $Cd_{0.6}Mn_{0.4}Te$ along the ingot.

X-ray analysis shows a cubic sphalerite structure of the grown crystals. The dependence between the lattice constants of the crystals and manganese concentration was obtained (see Fig. 3). The obtained values agree well with the existing data [13].

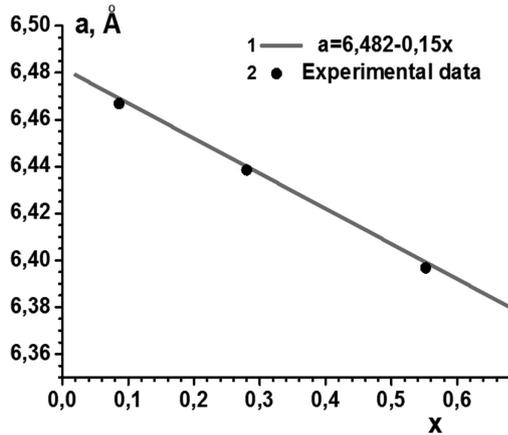

**Fig. 3.** Dependence between the manganese concentration and the lattice parameter of grown $Cd_{1-x}Mn_xTe$ crystals. The points correspond to the obtained experimental data.

The band-gap dependence of $Cd_{1-x}Mn_xTe:Fe^{2+}$ crystals on the solid solution composition was studied at high solid solution concentrations from $x=0.35$ to $x=0.57$. As shown in Fig. 4, the dependence has an almost linear character. The band gap increases with increasing solid solution concentration.

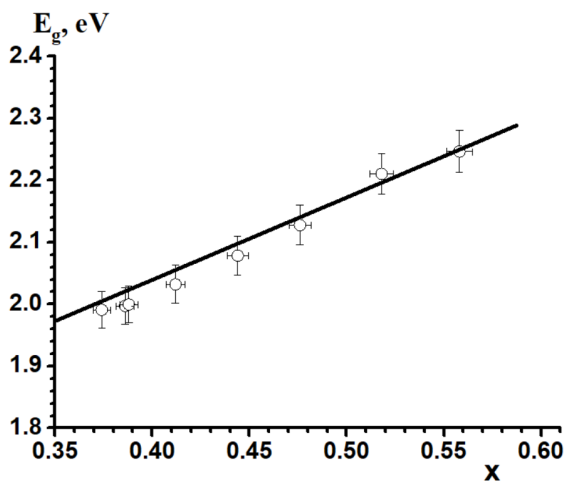

**Fig. 4.** Band gap concentration dependence of $Cd_{1-x}Mn_xTe:Fe^{2+}$ crystals.

Absorption and luminescence spectra of $Cd_{1-x}Mn_xTe:Fe^{2+}$ crystals in the mid-IR and visible ranges of the optical spectrum are studied. Fig. 5 shows the typical absorption band with a wavelength maximum near 3.6-3.8 micrometers, which corresponds to the intra-center electron transitions $^5E \to {}^5T_2$ for the $Fe^{2+}$ ions.

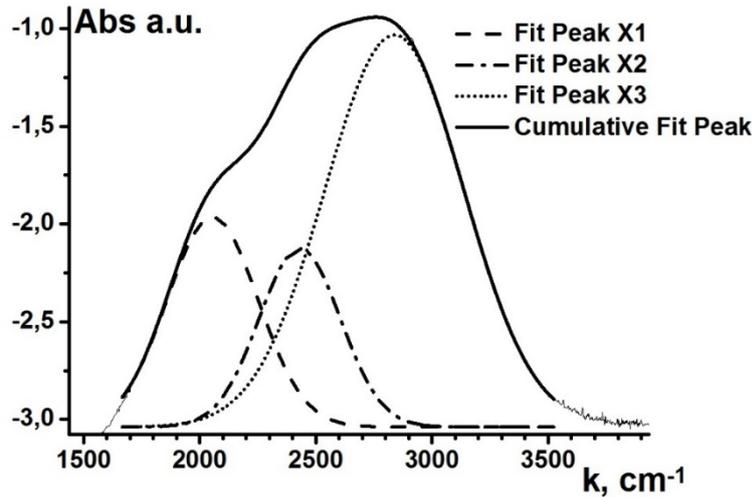

**Fig. 5.** Absorption bands of spectra of Fe2+ ions in the $Cd_{1-x}Mn_xTe:Fe^{2+}$ crystals at room temperature 293°K.

Processing of the transmission spectra of samples with different Mn contents allowed us to determine the position of the Jahn-Teller components (X1, X2, X3) for the optical absorption band caused by the $^5E \rightarrow {}^5T_2$ transition of $Fe^{2+}$ ions in $Cd_{1-x}Mn_xTe:Fe^{2+}$ crystals. The experimental data are collected in Table 2 and Table 3.

**Table 2.** Shifting of the maxima positions for the most long-wavelength (X3) Jahn-Teller component of the $Fe^{2+}$ ion absorption band for $Cd_{1-x}Mn_xTe:Fe^{2+}$ crystals with different solid solution compositions.

| Composition, $x$, at.% | Absorption maximum wavelength $\lambda_{ab}$, μm | Absorption "redshift" coefficient, $K_{ab}$, μm/at.% |
|---|---|---|
| <0.001 | 4.608 | — |
| 0.084 | 4.645 | 0.444 |
| 0.250 | 4.696 | 0.353 |
| 0.440 | 4.800 | 0.437 |
| 0.600 | 4.860 | 0.421 |
| 0.760 | 4.910 | 0.398 |

**Table 3.** Shifting of the maxima positions of the average absorption (by integrating all X1, X2 and X3 Jahn-Teller components) and emission spectra for $Cd_{1-x}Mn_xTe:Fe^{2+}$ crystals with different solid solution compositions.

| Composition, $x$, at.% | Absorption maximum wavelength $\lambda_{ab}$, μm | Absorption "redshift" coefficient, $K_{ab}$, μm/at.% | Emission maximum wavelength $\lambda_{em}$, μm | Emission "redshift" coefficient, $K_{em}$, μm/at.% | Difference, $\mid \Delta\lambda_{em}/\lambda_{em} - \Delta\lambda_{ab}/\lambda_{ab} \mid$ |
|---|---|---|---|---|---|
| <0.001 | 3.532 | — | 4.905 | — | — |
| 0.084 | 3.550 | 0.180 | 4.960 | 0.657 | 0.008 |
| 0.250 | 3.583 | 0.204 | 5.038 | 0.533 | 0.012 |
| 0.440 | 3.612 | 0.182 | 5.190 | 0.648 | 0.033 |
| 0.600 | 3.633 | 0.168 | 5.260 | 0.592 | 0.040 |
| 0.760 | 3.670 | 0.182 | 5.360 | 0.599 | 0.047 |

The concentration dependence of the absorption and luminescence band maxima of iron ions in $Cd_{1-x}Mn_xTe:Fe^{2+}$ solid solution crystals corresponds to the maximal shift in $\Delta\lambda_{ab} \sim 300$ nm and $\Delta\lambda_{em} \sim 400$ nm with increasing solid solution composition from $x = 0.1$ to $x = 0.76$.

The linear "redshift" law for the absorption band of the most longwavelength (X3) Jahn-Teller component has the form $\lambda_{ab}^{(X3)}(x) \approx K_{ab}^{(X3)} x + \lambda_{ab}^{(X3)}(x=0)$ with $K_{ab} \approx 0.411$ μm/at.%. The linear "redshift" law for the total absorption band integrated over all (X1, X2, X3) Jahn-Teller components has the form $\lambda_{ab}^{(total)}(x) \approx K_{ab}^{(total)} x + \lambda_{ab}^{(total)}(x=0)$ with $K_{ab}^{(total)} \approx 0.173$ μm/at.% and $K \sim K_{ab}^{(total)} / \lambda_{ab}^{(total)}(x=0) \approx 0.052$ /at.%. The latter value is close to the theoretical estimate $K \approx 0.063$.

The "redshift" law for the maximum of the emission spectra has a similar form: $\lambda_{em}(x) \approx K_{em} x + \lambda_{em}(x=0)$ with $K_{em} \approx 0.606$ μm/at.%. We have also checked the identity from Eq. (10) for the obtained experimental data (see the last column of Table 3). It works well for all concentrations of solid solution. Indeed, this fact confirms the previous theoretical conclusion that the maxima of the absorption and emission bands should shift in the same way to longer wavelengths with increasing $Cd_{1-x}Mn_xTe:Fe^{2+}$ solid solution concentration.

As we can see from the obtained data, further increasing the Mn concentration reduces the energy gap between the ground state and excited state of the active ion $Fe^{2+}$ and, in this way, shifts the luminescence and lasing band to the red side. We can state the offset of the Jahn-Teller component positions for optical absorption of $Fe^{2+}$ ions towards the longer-wavelength region with increasing Mn concentration, which indicates the extension of applicability limits for this material, e.g., as a mid-IR laser medium.

The linear sin-band redshift of absorption and emission spectra of transition metal ions (activators) considered in this work occurs for other crystals of chalcogenide solid solutions. In particular, a similar theoretical analysis [17] and previously performed experiments show that a redshift in the absorption and emission spectra also occurs for ternary crystals of solid solutions ZnMgSe [18] and ZnMnSe [19]. It is worth to note the following interesting feature – the replacement of the main cation in the solid solution matrix with a less electronegative element leads to a weakening of the average crystal field near the transition metal ion (chromium or iron), and, therefore, according to Eqs. (1)-(9), to a "redshift" of the absorption and emission spectra of the activator with increasing concentration of the solid solution, i.e., with increasing proportion of the replaced atoms. For example, this happens when cadmium cations (Polling's electronegativity $\chi = 1.69$) are replaced by less electronegative manganese cations ($\chi = 1.55$) in crystals of $Cd_{1-x}Mn_xTe:Fe^{2+}$ solid solution or when zinc cations ($\chi = 1.65$) are replaced by less electronegative magnesium cations ($\chi = 1.31$) or manganese cations ($\chi = 1.55$) in crystals of $Zn_{1-x}Mg_xSe:Cr^{2+}$ or $Zn_{1-x}Mn_xSe:Cr^{2+}$ solid solutions.

The magnitude of this redshift, i.e., the corresponding linear coefficient $K$ in Eqs. (4)-(8), depends on the relative change in the magnitude of the crystal field in the partial (binary) components of the solid solution, namely, from the ratio of the shift energy parameters $\Delta E^*_2 = \Delta E^*(x=1)$ and $\Delta E^*_1 = \Delta E^*(x=0)$ (see, for example, Eq. (9)). Ordinary, wavelength redshift is about ~20-50 nm per every 10% increment of cation replacement $x$ for the solid solution. If the electronegativities of the cations of ternary solid solution differ more and the difference between the partial energy shifts $\Delta E^*_2$ and $\Delta E^*_1$ is greater, then the linear redshift coefficient will be also the greater.

## 5. Conclusion

Doped semiconductor crystals of solid solution $Cd_{1-x}Mn_xTe:Fe^{2+}$ were grown by the high-pressure Bridgman method and investigated. A theoretical model is proposed that explains the observed shift of IR absorption and emission bands in the spectra of transition metal ions in solid solutions of semiconductor compounds. The model is based on the principle of additivity for semiconductor materials (similar to Vegard's law, etc.). It has been used for estimating the long-wavelength shift of absorption and luminescence bands in the spectra of semiconductor crystals $Cd_{1-x}Mn_xTe:Fe^{2+}$ with increasing solid solution concentration. The correlations between the solid solution composition and structural and band gap properties, as well as the maxima positions of the $Fe^{2+}$ ion absorption/emission spectra, were found. The obtained results can be used to predict the lasing range for $Cd_{1-x}Mn_xTe:Fe^{2+}$ active materials in all possible Mn concentrations.

**Acknowledgments**

This research was in part supported by the NATO Science for Peace and Security Programme (Project NATO SPS G5912).

**References**

[1] I.T. Sorokina, Crystalline Mid-Infrared Lasers, in *Solid-State Mid-Infrared Laser Sources*, Topics in Applied Physics, vol. 89, I.T. Sorokina, K.L. Vodopyanov, eds. (Springer, Berlin, Heidelberg, 2003), pp. 262-358. https://doi.org/10.1007/3-540-36491-9_7

[2] S.B. Mirov, I.S. Moskalev, S. Vasilyev, V. Smolski, V.V. Fedorov, D. Martyshkin, J. Peppers, M. Mirov, A. Dergachev, V. Gapontsev, "Frontiers of Mid-IR Lasers Based on Transition Metal Doped Chalcogenides", IEEE Journal of Selected Topics in Quantum Electronics 24 (2018) 1601829. https://doi.org/10.1109/JSTQE.2018.2808284

[3] A.V Podlipensky, V.G Shcherbitsky, M.I Demchuk, N.V. Kuleshov, V.I Levchenko, V.N Yakimovich, S Girard, R Moncorge, "$Cr^{2+}$:$Cd_{0.55}Mn_{0.45}Te$ crystal as a new saturable absorber for 2 µm lasers", Optics Communications 192 (2001) 65-68. https://doi.org/10.1016/S0030-4018(01)01025-2

[4] M. Mond, D. Albrecht, E. Heumann, G. Huber, S. Kück, V.I. Levchenko, V.N. Yakimovich, V.G. Shcherbitsky, V.E. Kisel, N.V. Kuleshov, M. Rattunde, J. Schmitz, R. Kiefer, J. Wagner, "1.9-m and 2.0-m laser diode pumping of $Cr^{2+}$:ZnSe and $Cr^{2+}$:CdMnTe", Opt. Lett. 27 (2002) 1034-1036. https://doi.org/10.1364/OL.27.001034

[5] S.B. Trivedi, C.C. Wang, S. Kutcher, U. Hommerich, W. Palosz, "Crystal Growth Technology of Binary and Ternary II-VI Semiconductors for Photonic Applications", Journal of Crystal Growth 310 (2008) 1099-1106.

[6] V.V. Fedorov, W. Mallory, S.B. Mirov, U. Hömmerich, S.B. Trivedi, W. Palosz, "Iron-doped $Cd_xMn_{1-x}Te$ crystals for mid-IR room temperature lasers", Journal of Crystal Growth 310 (2008) 4438-4442. https://doi.org/10.1016/j.jcrysgro.2008.07.110

[7] H. Jelinkova, J. Sulc, M. Jelinek, M. Nemec, D. Vyhlidal, M. E. Doroshenko, V. Osiko, N. Kovalenko, A. Gerasimenko, "Iron doped $Cd_xMn_{1-x}Te$ Active Material for Mid-IR lasers", in Proc. of Advanced Solid State Lasers 2015 (Berlin Germany, 4-9 October 2015), paper ATu2A.16. https://doi.org/10.1364/ASSL.2015.ATu2A.16

[8] H. Jelinkova, M.E. Doroshenko, M. Jelinek, J. Sulc, M. Nemec, V.V. Osiko, N.O. Kovalenko, I.S. Terzin, "Fe:CdMnTe active material spectroscopic properties and laser generation around 5 µm", in Proc. SPIE 10082, Solid State Lasers XXVI: Technology and Devices (San Francisco, California, 2017) 100820F. https://doi.org/10.1117/12.2252317

[9] A. Mycielski, A. Burger, M. Sowinska, M. Groza, A. Szadkowski, P. Wojnar, B. Witkowska, W. Kaliszek, P. Siffert, "Is the (Cd,Mn)Te Crystals a Perspective Material for X-Ray and γ-Ray Detectors", J. Phys. Stat. Sol. (C) 2 (2005) 1578-1585. https://doi.org/10.1002/pssc.200460838

[10] A. Hossain, Y. Cui, A. Bolotnikov, G. Camarda, G. Yang, K-H. Kim, R. Gul, L. Xu, L. Li, A. Mycielski, R.B. James, "Cadmium Manganese Telluride ($Cd_{1-x}Mn_xTe$): A potential material for room-temperature radiation detectors", report No. BNL-93721-2010CP (Brookhaven National Laboratory, USA, 2010). https://digital.library.unt.edu/ark:/67531/metadc839980/

[11] L. Luan, L. Gao, H. Lv, P. Yu, T. Wang, Y. He, D. Zheng, "Analyses of crystal growth, optical, electrical, thermal and mechanical properties of an excellent detector grade $Cd_{0.9}Mn_{0.1}Te$:V crystal", Sci. Rep. 10 (2020) 2749. https://doi.org/10.1038/s41598-020-59612-0

[12] R. Triboulet, G. Didier, "Grows and characterization of $Cd_{1-x}Mn_xTe$ and MnTe crystals; contribution to the CdTe-MnTe pseudo-binary phase diagram determination", Journal of Crystal Growth 52 (1981) 614-618. https://doi.org/10.1016/0022-0248(81)90350-X

[13] Y. Cui, A. Bolotnikov, A. Hossain, G. Camarda, A. Mycielski, G. Yang, D. Kochanowska, M. Witkowska-Baran, R.B. James, "CdMnTe crystals for x-ray and gamma-ray detection", in Proc. SPIE 7079, Hard X-Ray, Gamma-Ray, and Neutron Detector Physics X, 70790N (4 September 2008). https://doi.org/10.1117/12.793366

[14] A. Zunger, "Electronic Structure of 3d Transition-Atom Impurities in Semiconductors", in Solid State Physics Vol. 39, Ed. by H. Ehrenreich and D. Turnbull (Academic Press, 1986). https://doi.org/10.1016/S0081-1947(08)60371-9

[15] T. P. Surkova, M. Godlewski, K. Swiatek, P. Kaczor, A.Polimeni, L. Eaves, W. Giriat, "Intra-shell transitions of 3D metal ions (Fe, Co, Ni) in II–VI wide-gap semiconductor alloys", Phys. B (Amsterdam, Neth.) 848 (1999) 273-274. https://doi.org/10.1016/S0921-4526(99)00519-0

[16] I.B. Bersuker, Electronic Structure and Properties of Coordination Compounds. Introduction to Theory, 2nd ed. (John Wiley & Sons, Hoboken, New Jersey, 2010).

[17] S.V. Naydenov, Shift of the Infrared Absorption and Emission Spectra of Transition Metal Ions in Solid Solutions of Semiconductor Compounds, Tech. Phys. Lett. 47 (2021) 613-616. https://doi.org/10.1134/S1063785021060249

[18] M.E. Doroshenko, V.V. Osiko, H. Jelínková, M. Jelínek, M. Němec, J. Šulc, N.O. Kovalenko, A.S. Gerasimenko, V.M. Puzikov, Spectroscopic and laser properties of $Cr^{2+}$ ions in $Zn_{1-x}Mg_xSe$ solid solutions, Optical Materials 47 (2015) 185-189. https://doi.org/10.1016/j.optmat.2015.05.015

[19] M.E. Doroshenko, H. Jelinkova, A. Riha, K. Pierpoint, Spectroscopic properties of $Cr^{2+}$ ions in $Zn_{1-x}Mn_xSe$ solid solutions, Materials Today Communications 37 (2023), 107048. https://doi.org/10.1016/j.mtcomm.2023.107048